\documentclass[twocolumn,showpacs,aps,prl,superscriptaddress]{revtex4}

\usepackage{multirow}
\usepackage{graphicx}
\usepackage{dcolumn}
\usepackage{amsmath}
\usepackage{epsfig}

\RequirePackage{xspace}

\usepackage{relsize}
\def\babar{\mbox{\slshape B\kern-0.1em{\smaller A}\kern-0.1em
    B\kern-0.1em{\smaller A\kern-0.2em R}}}

\def\epem       {\ensuremath{e^+e^-}\xspace}

\def\mumu       {\ensuremath{\mu^+\mu^-}\xspace}

\def\ellell     {\ensuremath{\ell^+ \ell^-}\xspace}

\def\piz   {\ensuremath{\pi^0}\xspace}

\def\pipi  {\ensuremath{\pi^+\pi^-}\xspace}

\def\Kbar  {\kern 0.2em\overline{\kern -0.2em K}{}\xspace}

\def\Kz    {\ensuremath{K^0}\xspace}
\def\Kzb   {\ensuremath{\Kbar^0}\xspace}
\def\KzKzb {\ensuremath{\Kz \kern -0.16em \Kzb}\xspace}
\def\Kp    {\ensuremath{K^+}\xspace}
\def\Km    {\ensuremath{K^-}\xspace}

\def\KpKm  {\ensuremath{\Kp \kern -0.16em \Km}\xspace}

\def\Dbar    {\kern 0.2em\overline{\kern -0.2em D}{}\xspace}

\def\Dz      {\ensuremath{D^0}\xspace}
\def\Dzb     {\ensuremath{\Dbar^0}\xspace}
\def\DzDzb   {\ensuremath{\Dz {\kern -0.16em \Dzb}}\xspace}
\def\Dp      {\ensuremath{D^+}\xspace}
\def\Dm      {\ensuremath{D^-}\xspace}

\def\DpDm    {\ensuremath{\Dp {\kern -0.16em \Dm}}\xspace}

\def\Bbar    {\kern 0.18em\overline{\kern -0.18em B}{}\xspace}

\def\BB      {\ensuremath{B\Bbar}\xspace} 
\def\Bz      {\ensuremath{B^0}\xspace}
\def\Bzb     {\ensuremath{\Bbar^0}\xspace}
\def\BzBzb   {\ensuremath{\Bz {\kern -0.16em \Bzb}}\xspace}
\def\Bu      {\ensuremath{B^+}\xspace}
\def\Bub     {\ensuremath{B^-}\xspace}

\def\BpBm    {\ensuremath{\Bu {\kern -0.16em \Bub}}\xspace}

\def\BorBbar    {\kern 0.18em\optbar{\kern -0.18em B}{}\xspace}
\def\DorDbar    {\kern 0.18em\optbar{\kern -0.18em D}{}\xspace}
\def\KorKbar    {\kern 0.18em\optbar{\kern -0.18em K}{}\xspace}

\mathchardef\Upsilon="7107
\def\Y#1S{\ensuremath{\Upsilon{(#1S)}}\xspace}% no space before {...}!
\def\OneS  {\Y1S}
\def\TwoS  {\Y2S}
\def\ThreeS{\Y3S}
\def\FourS {\Y4S}

\mathchardef\Deltares="7101
\mathchardef\Xi="7104
\mathchardef\Lambda="7103
\mathchardef\Sigma="7106
\mathchardef\Omega="710A

\def\Deltabar{\kern 0.25em\overline{\kern -0.25em \Deltares}{}\xspace}
\def\Lbar{\kern 0.2em\overline{\kern -0.2em\Lambda\kern 0.05em}\kern-0.05em{}\xspace}
\def\Sigbar{\kern 0.2em\overline{\kern -0.2em \Sigma}{}\xspace}
\def\Xibar{\kern 0.2em\overline{\kern -0.2em \Xi}{}\xspace}
\def\Obar{\kern 0.2em\overline{\kern -0.2em \Omega}{}\xspace}
\def\Nbar{\kern 0.2em\overline{\kern -0.2em N}{}\xspace}
\def\Xb{\kern 0.2em\overline{\kern -0.2em X}{}\xspace}

\def\BR         {{\ensuremath{\cal B}\xspace}}

\newcommand{\tev}{\ensuremath{\mathrm{\,Te\kern -0.1em V}}\xspace}
\newcommand{\gev}{\ensuremath{\mathrm{\,Ge\kern -0.1em V}}\xspace}
\newcommand{\mev}{\ensuremath{\mathrm{\,Me\kern -0.1em V}}\xspace}
\newcommand{\kev}{\ensuremath{\mathrm{\,ke\kern -0.1em V}}\xspace}
\newcommand{\ev}{\ensuremath{\mathrm{\,e\kern -0.1em V}}\xspace}
\newcommand{\gevc}{\ensuremath{{\mathrm{\,Ge\kern -0.1em V\!/}c}}\xspace}
\newcommand{\mevc}{\ensuremath{{\mathrm{\,Me\kern -0.1em V\!/}c}}\xspace}
\newcommand{\gevcc}{\ensuremath{{\mathrm{\,Ge\kern -0.1em V\!/}c^2}}\xspace}
\newcommand{\mevcc}{\ensuremath{{\mathrm{\,Me\kern -0.1em V\!/}c^2}}\xspace}
\def\invfb   {\ensuremath{\mbox{\,fb}^{-1}}\xspace}

\def\mus  {\ensuremath{\rm \,\mus}\xspace}

\def\mus        {\ensuremath{\,\mu{\rm s}}\xspace}    %% microsecond
\def\to                 {\ensuremath{\rightarrow}\xspace}

\def\pep2{PEP-II}

\def\gsim{{~\raise.15em\hbox{$>$}\kern-.85em
          \lower.35em\hbox{$\sim$}~}\xspace}
\def\lsim{{~\raise.15em\hbox{$<$}\kern-.85em
          \lower.35em\hbox{$\sim$}~}\xspace}

\xspace

\def\evtgen     {\mbox{\tt EvtGen}\xspace}

\def\geant      {\mbox{\tt GEANT}\xspace}

\def\jetset74   {\mbox{\tt Jetset \hspace{-0.5em}7.\hspace{-0.2em}4}\xspace}
\newcommand{\BABARPubYear}    {11}
\newcommand{\BABARPubNumber}  {018}

\newcommand{\SLACPubNumber} {14546}

\def\figurebox#1#2#3{%
    \def\arg{#3}%
    \ifx\arg\empty
    {\hfill\vbox{\hsize#2\hrule\hbox to #2{\vrule\hfill\vbox to #1{\hsize#2\vfill}\vrule}\hrule}\hfill}%
    \else
    {\hfill\epsfbox{#3}\hfill}%
    \fi}

\begin{document}

\preprint{\babar-PUB-\BABARPubYear/\BABARPubNumber} 
\preprint{SLAC-PUB-\SLACPubNumber} 

\begin{flushleft}
\babar-PUB-\BABARPubYear/\BABARPubNumber\\
SLAC-PUB-\SLACPubNumber\\
\end{flushleft}

\title{
{\large \bf  \boldmath
Study of $\Upsilon(3S,2S)\to\eta\OneS$ and $\Upsilon(3S,2S)\to\pipi\OneS$ hadronic transitions
}}

%% author list as of 01-Jul-2011 (387 authors)
%
\author{J.~P.~Lees}
\author{V.~Poireau}
\author{V.~Tisserand}
\affiliation{Laboratoire d'Annecy-le-Vieux de Physique des Particules (LAPP), Universit\'e de Savoie, CNRS/IN2P3,  F-74941 Annecy-Le-Vieux, France}
\author{J.~Garra~Tico}
\author{E.~Grauges}
\affiliation{Universitat de Barcelona, Facultat de Fisica, Departament ECM, E-08028 Barcelona, Spain }
\author{M.~Martinelli$^{ab}$}
\author{D.~A.~Milanes$^{a}$}
\author{A.~Palano$^{ab}$ }
\author{M.~Pappagallo$^{ab}$ }
\affiliation{INFN Sezione di Bari$^{a}$; Dipartimento di Fisica, Universit\`a di Bari$^{b}$, I-70126 Bari, Italy }
\author{G.~Eigen}
\author{B.~Stugu}
\affiliation{University of Bergen, Institute of Physics, N-5007 Bergen, Norway }
% ...
\collaboration{The \babar\ Collaboration}
\noaffiliation

\date{August 30, 2011}%It is always \today, today, but you may specify any date with \date.

\begin{abstract}
We  study the $\Upsilon(3S,2S)\to\eta\OneS$ and $\Upsilon(3S,2S)\to\pipi\OneS$ transitions with 122$\times10^{6}$ $\ThreeS$ and 100$\times10^{6}$ $\TwoS$ mesons collected by the \babar\ detector at the \pep2\ asymmetric-energy $\epem$ collider. We measure $\displaystyle{\cal B}[\TwoS\to\eta\OneS]=(2.39\pm0.31{\rm (stat.)}\pm0.14{\rm (syst.)})\times10^{-4}$ and $\displaystyle\Gamma[\TwoS\to\eta\OneS]/\Gamma[\TwoS\to\pipi\OneS]=(1.35\pm0.17{\rm(stat.)}\pm0.08{\rm (syst.)})\times10^{-3}$.  We find no evidence for $\ThreeS\to\eta\OneS$ and obtain $\displaystyle{\cal B}[\ThreeS\to\eta\OneS]<1.0\times10^{-4}$ and $\displaystyle\Gamma[\ThreeS\to\eta\OneS]/\Gamma[\ThreeS\to\pipi\OneS]<2.3\times10^{-3}$ as upper limits at the 90$\%$ confidence level. We also provide improved measurements of the $\Upsilon(2S) - \Upsilon(1S)$ and $\Upsilon(3S) - \Upsilon(1S)$ mass differences, $562.170\pm0.007{\rm(stat.)}\pm0.088{\rm(syst.)}\mevcc$ and $893.813\pm0.015{\rm(stat.)}\pm0.107{\rm(syst.)}\mevcc$, respectively.
\end{abstract}

% PACS, the Physics and Astronomy Classification Scheme.
\maketitle

The QCD multipole expansion (QCDME) model~\cite{ref:Kuang} describes hadronic transitions between heavy quarkonia.
Despite its success for hadronic transitions in charmonium,  this model has
limits in explaining all hadronic transitions in the bottomonium spectrum.  
The QCDME predicts the suppression of the transitions between bottomonia via a $\eta$ meson with respect to those via a dipion, the former being associated with the spin-flip effects of the $b$ quark. 
The $\FourS\to\eta\OneS$ and $\TwoS\to\eta\OneS$ transitions have been observed at rates significantly different from the predicted values~\cite{ref:CLEO_eta,ref:BaBar_eta}. The measured width $\Gamma[\TwoS\to\eta\OneS]$ is smaller than predicted, while $\Gamma[\FourS\to\eta\OneS]$ is larger than $\Gamma[\FourS\to\pipi\OneS]$, although it was expected to be suppressed in analogy with decays of the lower-mass $\Upsilon$ resonances. Some suggest that the latter result could be related to above-\BB threshold effects~\cite{ref:Simonov, ref:Meng}. The $\ThreeS\to\eta\OneS$ transitions have not been observed~\cite{ref:CLEO_eta}.
Precise measurements of the transitions between bottomonia via a $\eta$ meson, as well as their rate with respect to the dipion transitions could shed light on the chromomagnetic moment of the $b$ quark.

In this paper we study the transitions $\Upsilon(nS)\to\eta\OneS$ and $\Upsilon(nS)\to\pipi\OneS$ with $n=3,2$ and measure the ratios of partial widths $\Gamma[\Upsilon(nS)\to\eta\OneS]/\Gamma[\Upsilon(nS)\to\pipi\OneS]$.
The transitions are studied for events in which the $\OneS$ decays to either $\mumu$ or $\epem$. The $\eta$ meson is reconstructed from its  $\gamma\gamma$ and $\pipi\pi^0$ decay modes, where the $\pi^0$ decays to $\gamma\gamma$. The analysis thus considers
the final states $\pipi\gamma\gamma \ellell$, $\gamma\gamma \ellell$ and $\pipi \ellell$, where $\ell =$ $e$ or $\mu$.

We analyze \babar\ data samples consisting of $(121.8\pm1.2)\times10^6$ $\ThreeS$ and $(98.6\pm0.9)\times10^6$ $\TwoS$ mesons. These correspond to integrated luminosities of $28.0$ fb$^{-1}$ and $13.6$ fb$^{-1}$, respectively. 
We use 2.6~\invfb\ collected 30~MeV below the $\ThreeS$ resonance, and 1.4~\invfb\ collected 30~MeV below the $\TwoS$ resonance (``off-peak'' samples) for background studies.

The \babar\ detector is described in detail elsewhere~\cite{ref:babar1,ref:babar2}. We briefly mention the features relevant to this analysis.
Charged-particle momenta are measured in  a five-layer double-sided silicon vertex tracker (SVT) and a 40-layer central drift chamber (DCH), both embedded in a 1.5-T axial magnetic field. Charged-particle identification is based on specific energy loss in the SVT and the DCH and on measurements of the photons produced in the fused-silica bars of a ring-imaging Cherenkov detector. A CsI(Tl) electromagnetic calorimeter (EMC) is used to detect and identify photons and to identify electrons, while muons are identified in the instrumented flux return of the magnet (IFR).

Monte Carlo (MC) simulated events, used for efficiency determination and selection optimization, are generated with \evtgen~\cite{ref:evtgen}; \geant4~\cite{ref:geant} is used to simulate the detector response. The variations of conditions and beam backgrounds are taken into account in the simulation. 
The simulated events are then analyzed in the same manner as data.
Large MC samples simulating inclusive $\Upsilon(3S)$ and $\Upsilon(2S)$ decays including all known and predicted transitions and continuum $\epem\to\epem(\gamma)$ and $\epem\to\mumu(\gamma)$ processes are used to characterize the backgrounds. Background from continuum quark production is negligible.
In the MC signal samples, the distribution of generated dilepton decays incorporates the $\OneS$ polarization. Dipion transitions are modeled according to the matrix elements measured by CLEO~\cite{ref:CLEO_dipion}. The angular distribution in $\Upsilon(3S,2S)\to\eta\OneS$ processes is generated as a vector decaying to a pseudoscalar and a vector. The $\eta\to\pipi\pi^0$ decays are modeled according to the known Dalitz plot parameters~\cite{ref:PDG2010}.
Final state radiation effects are described by \texttt{PHOTOS}~\cite{ref:photos}. 

Events of interest contain two oppositely charged particles, identified as either electrons or muons. 
A fit constrains them to originate from a common vertex and to have invariant mass $M_{\ell\ell}$ equal to the known $\OneS$ mass~\cite{ref:PDG2010}. The fit must yield a $\chi^2$ probability $>10^{-5}$.
Muon identification is based on the energy deposited in the EMC, and the number of coordinates and interaction lengths traversed in the IFR. Electron identification is based on specific energy loss in the SVT and DCH combined with energy deposition in the EMC. Bremsstrahlung energy loss is partially recovered by an algorithm combining the energy of an electron candidate with the energies of nearby photons.

Besides the lepton pair, we require a pair of oppositely charged tracks not identified as electrons and/or two neutral particles identified as photon candidates.
Events with additional charged tracks are rejected. 
A fit constrains all final state particles to originate from a common vertex, to have a total energy equal to the sum of the beam energies, and an invariant mass equal to the $\Upsilon(3S)$ or $\Upsilon(2S)$ mass~\cite{ref:PDG2010}. The fit must yield a $\chi^2$ probability $>10^{-5}$.

A trigger-level prescaling of Bhabha scattering events, whose signature is given by two electrons of large invariant mass and no additional charged track of transverse momentum $>250\mevc$, causes the efficiency for the final states containing electrons to be smaller than for final states with muons. The di-electron efficiency drops to $\sim0$ for $\TwoS$  transitions in all final states considered, and it is $<3\%$ for  $\ThreeS\to\eta\OneS$ in the $\gamma\gamma\epem$ final state. These final states are not considered further.

The event selection criteria have been optimized separately for each final state. The background contributions have been studied using MC samples of inclusive $\Upsilon(nS)$ decays and of $\epem(\gamma)$ and $\mumu(\gamma)$ events. 
The MC background yield has been compared to real background yield from data and found to be compatible with it within the uncertainties. Also it has been verified that the distributions of all the discriminating variables are well-described by the MC background.

No further selection is applied for the $\pipi \ellell$ final states, while the additional requirements summarized in Table~\ref{tab:selection} are needed for the other final states. 
To select the $\pipi\gamma\gamma\ellell$ final states we require that the two-photon invariant mass $M_{\gamma\gamma}$ be compatible with the $\pi^0$ mass. 
Background events are rejected by applying selection criteria to the opening angle between the two pions, calculated in the $\epem$ center-of-mass (CM) frame ($\theta_{\pi\pi}^*$), and also to the invariant mass of the dipion candidate calculated assuming the electron mass hypothesis ($M_{conv}$). In particular for the $\pipi\gamma\gamma\epem$ states, $M_{conv}>30$~MeV suppresses events in which a photon converts in the detector material and the electrons are reconstructed as pions.
In the $\Upsilon(3S)\to\eta\OneS$ final state, cross-feed from $\Upsilon(3S,2S)\to\pipi\Upsilon(2S,1S)$ transitions is suppressed by vetoing events with  $\Delta M_{\pi\pi}\equiv M_{\pi\pi\ell\ell}-M_{\ell\ell}$ compatible with any of the known mass differences between narrow $\Upsilon$ resonances. 
In the $\gamma\gamma\mumu$ final states, the backgrounds due to the radiative transitions $\Upsilon(nS)\to\gamma\chi_{bJ}(2P,1P)$ with $\chi_{bJ}(2P,1P)\to\gamma\OneS$ are rejected by vetoing events where either photon energy calculated in the CM frame ($E_{\gamma1,\gamma2}^*$) is compatible with any of those transitions.
The background from $\mumu(\gamma)$ events is reduced by requirements on the opening angle between the two photons ($\theta_{\gamma\gamma}^*$), and on the momentum of each lepton ($p_{\ell^\pm}^*$) in the CM frame.

\begin{table}[htdp]
\caption{Additional requirements applied to select $\pipi\gamma\gamma \ellell$  and $\gamma\gamma \ellell$ final states. Masses are expressed in $\mevcc$, energies in $\gev$, and momenta in $\gevc$.}
\begin{center}
\begin{tabular}{ccc}
\hline
\hline
$\TwoS\to\eta\OneS$ & \multicolumn{2}{c}{$\ThreeS\to\eta\OneS$}\\
\hline
$\pipi\gamma\gamma\mumu$ & $\pipi\gamma\gamma\mumu$ & $\pipi\gamma\gamma\epem$ \\
\hline
$M_{conv}<310$~ & ~$M_{conv}<280$ & $ 30<M_{conv}<280$\\
$90<M_{\gamma\gamma}<180$~ &~ $100<M_{\gamma\gamma}<170$~  & ~$90<M_{\gamma\gamma}<150$\\
$\cos\theta_{\pi\pi}^*<0$ & & \\
& \multicolumn{2}{c}{ ~$(400<\Delta M_{\pi\pi}<550)\cup(\Delta M_{\pi\pi}>580)$}~\\
\hline
$\gamma\gamma\mumu$ & \multicolumn{2}{c}{$\gamma\gamma\mumu$} \\
\hline
$200<E_{\gamma1,\gamma2}^*<350$~ & \multicolumn{2}{c}{$(120<E_{\gamma1}^*<360)\cup(490<E_{\gamma1}^*<660)$}\\
 &\multicolumn{2}{c}{$(130<E_{\gamma2}^*<370)\cup(470<E_{\gamma2}^*<700)$}\\
$\cos\theta_{\gamma\gamma}^*<-0.86$~ & \multicolumn{2}{c}{$-0.45<\cos\theta_{\gamma\gamma}^*<0.22$}\\
$4.60<p_{\ell^\pm}^*<4.85$ & \multicolumn{2}{c}{$4.4<p_{\ell^+}^*<5.0$; $4.4<p_{\ell^-}^*<5.1$}\\
\hline
\hline
\end{tabular}
\end{center}
\label{tab:selection}
\end{table}

\begin{figure}[htdp]
\begin{center}
\begin{tabular}{lll}
\includegraphics[scale=0.23]{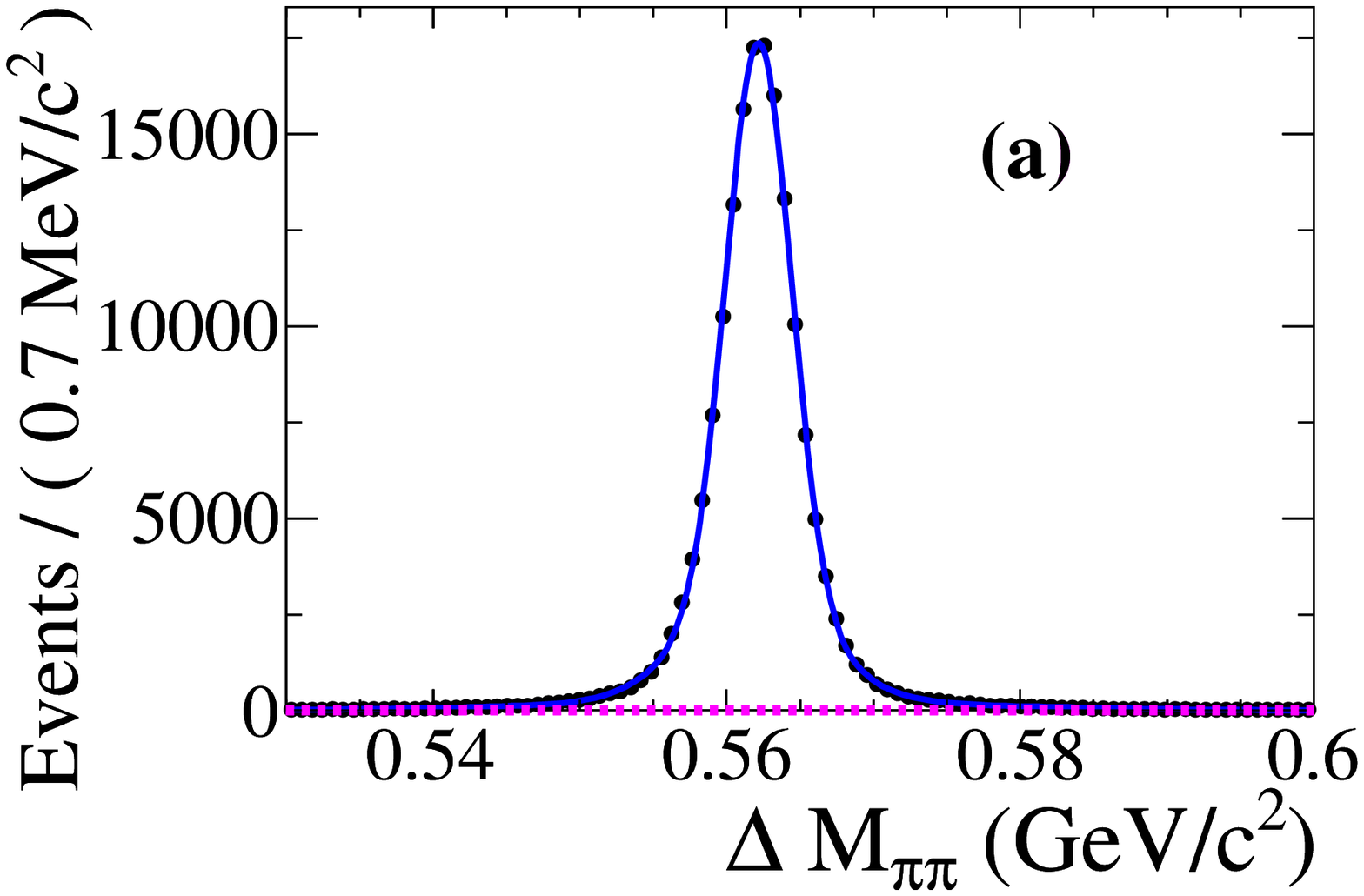}&
\includegraphics[scale=0.23]{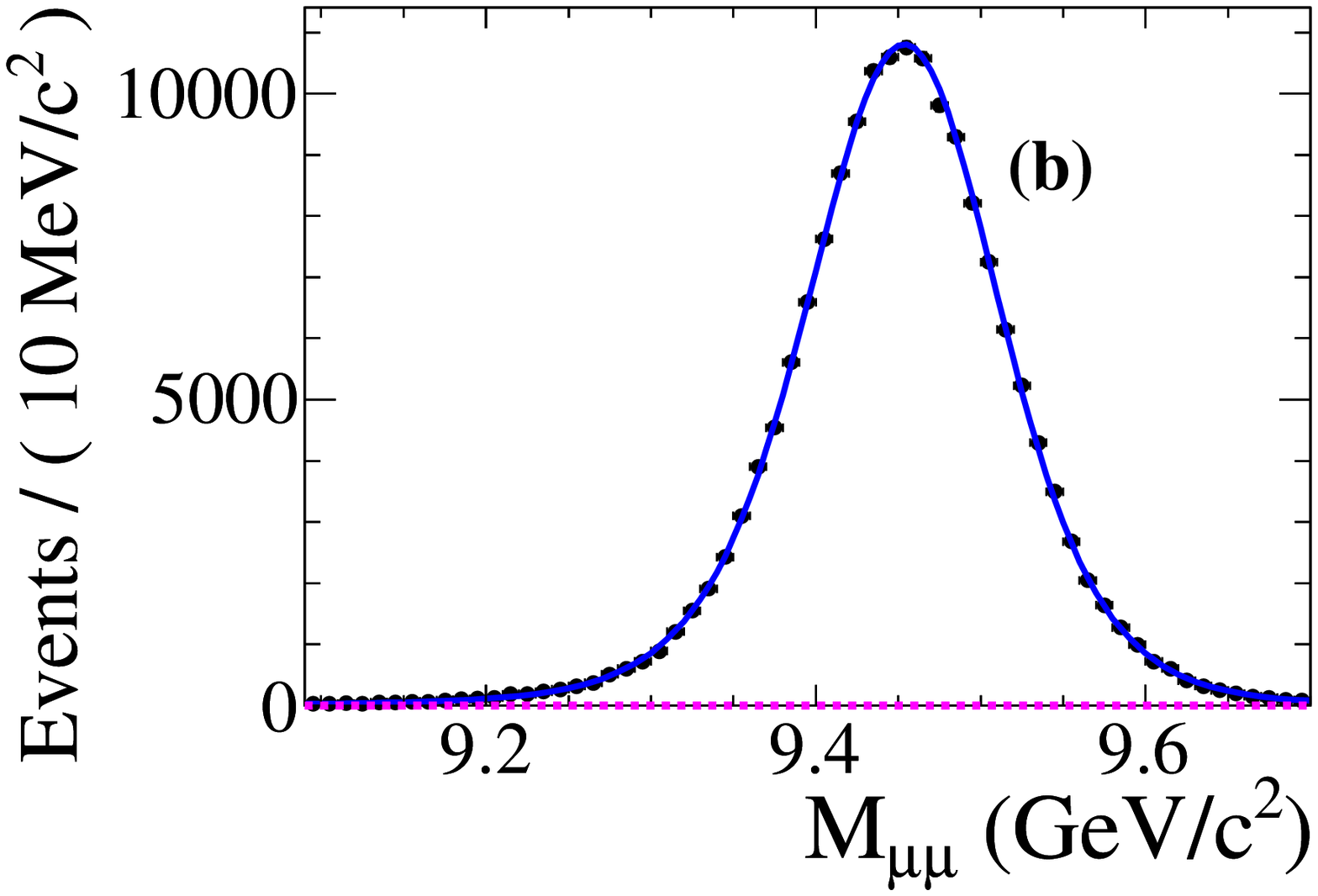}\\
\includegraphics[scale=0.23]{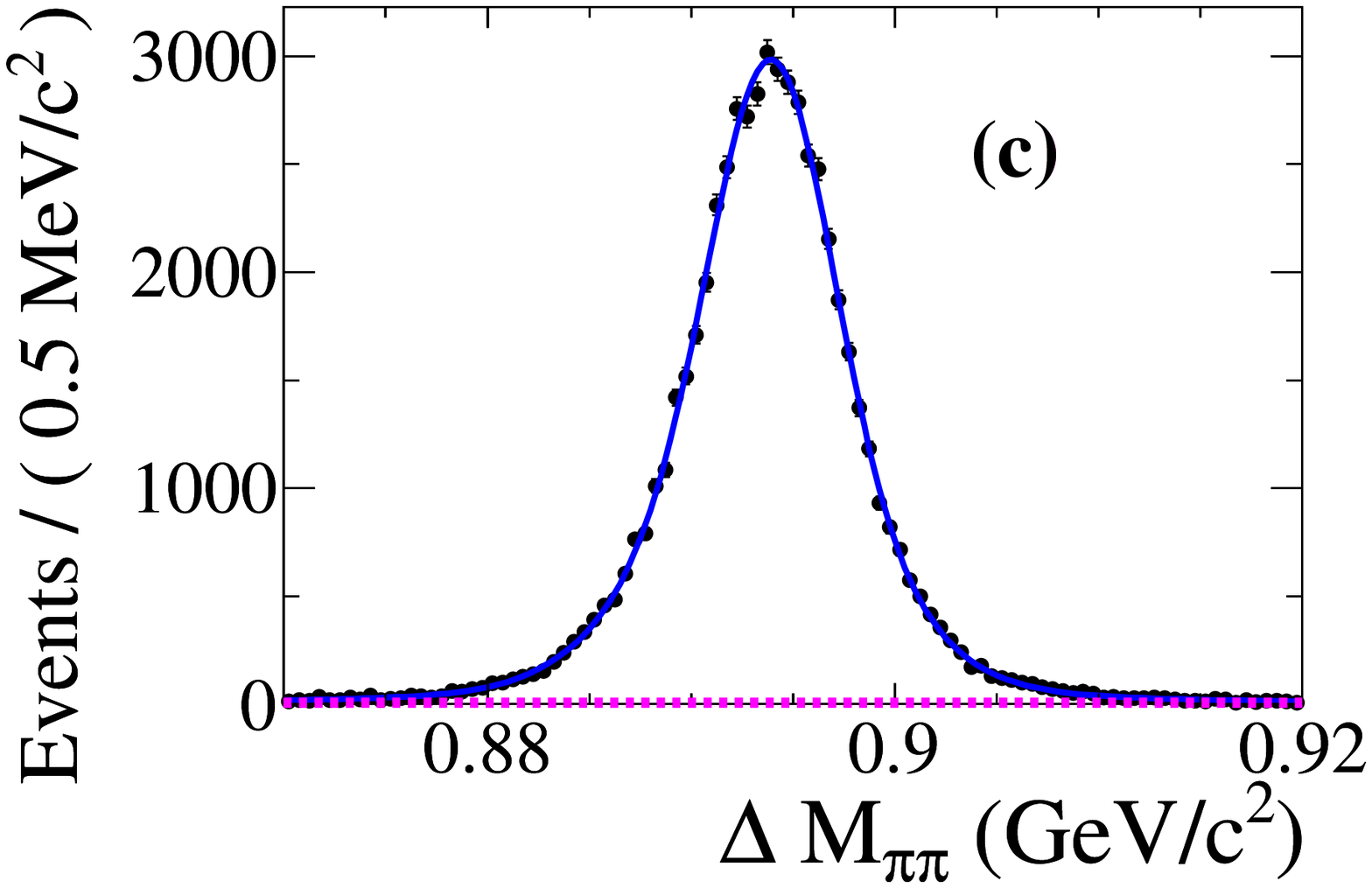}&
\includegraphics[scale=0.23]{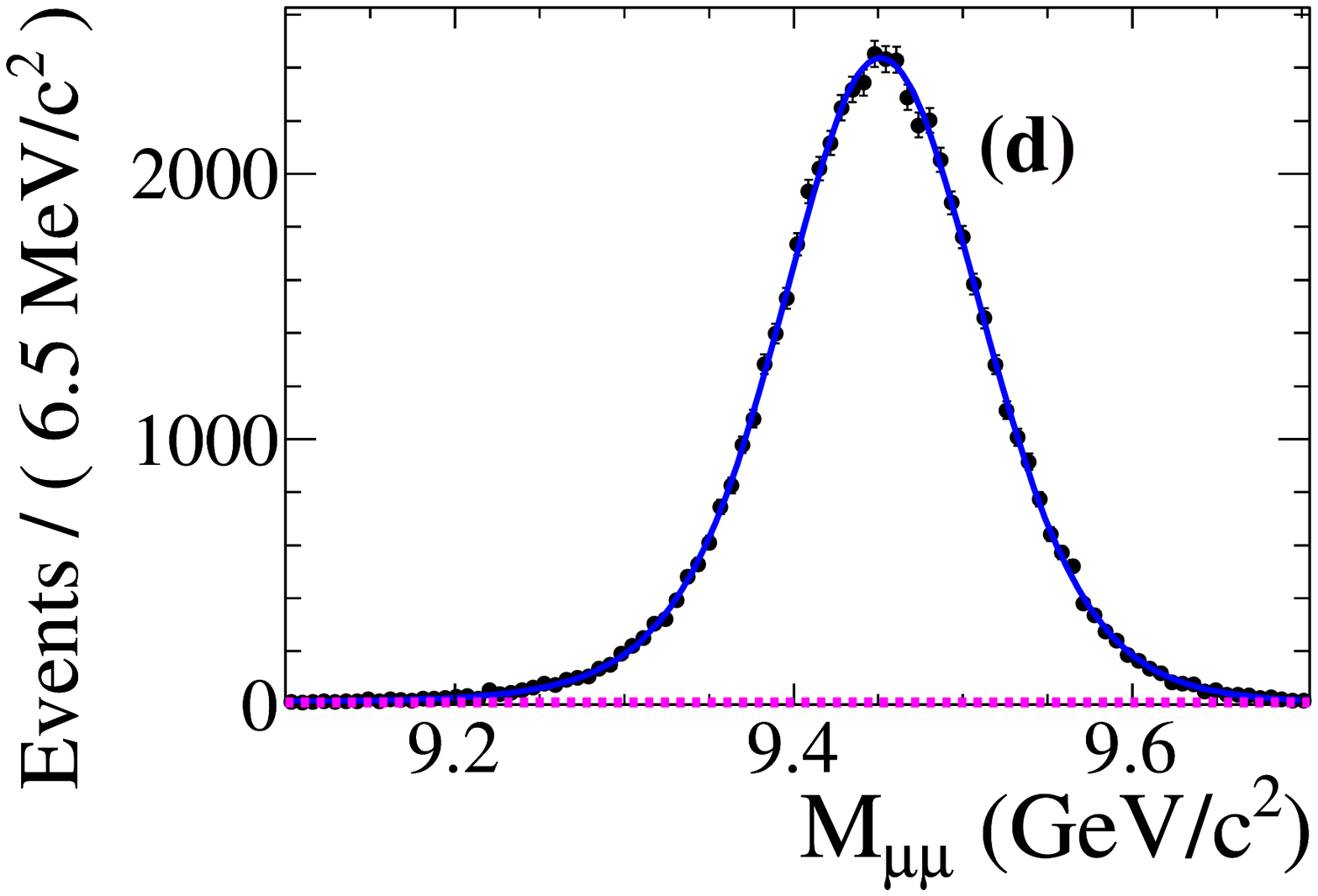}\\
\includegraphics[scale=0.23]{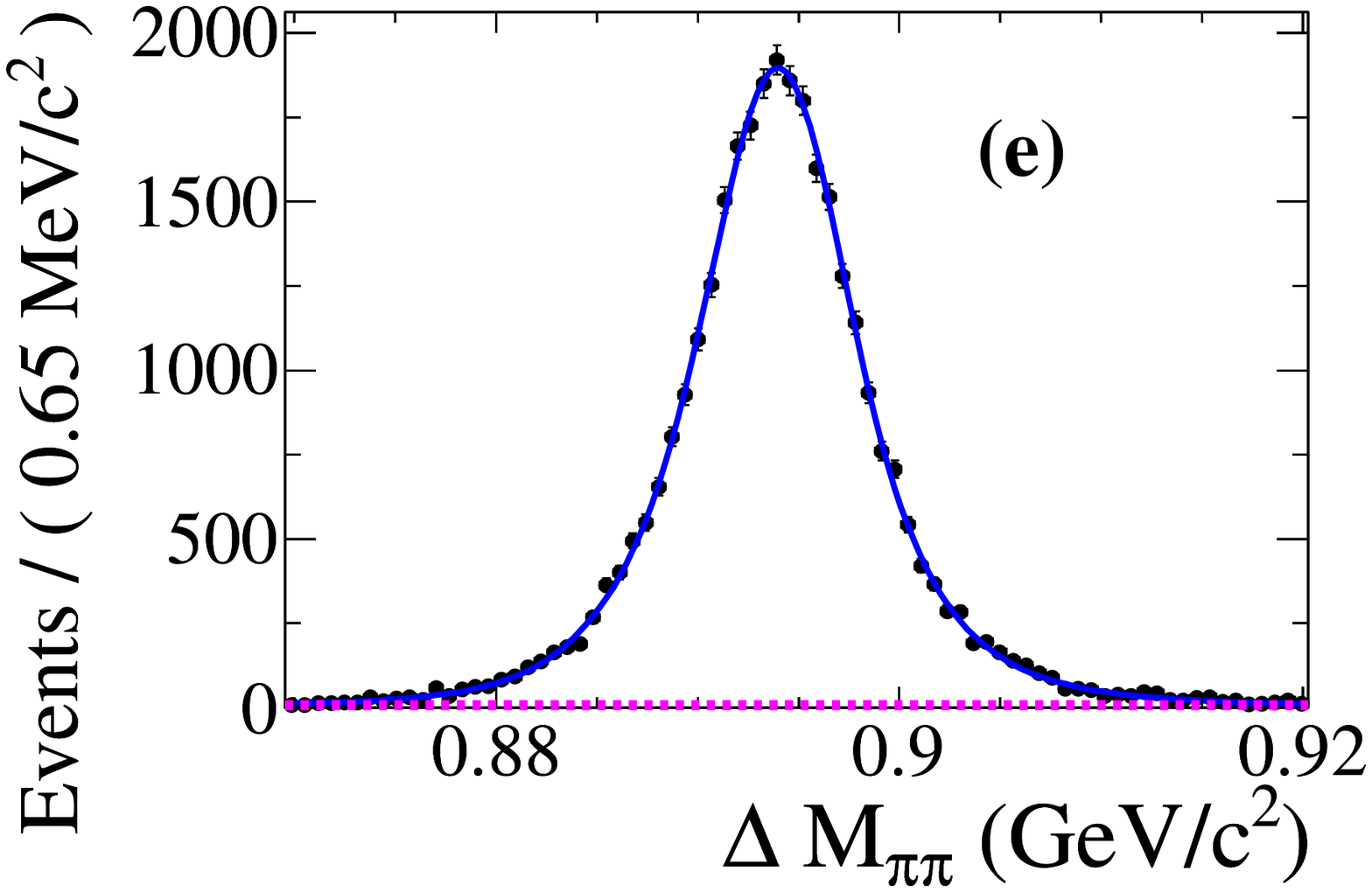}&
\includegraphics[scale=0.23]{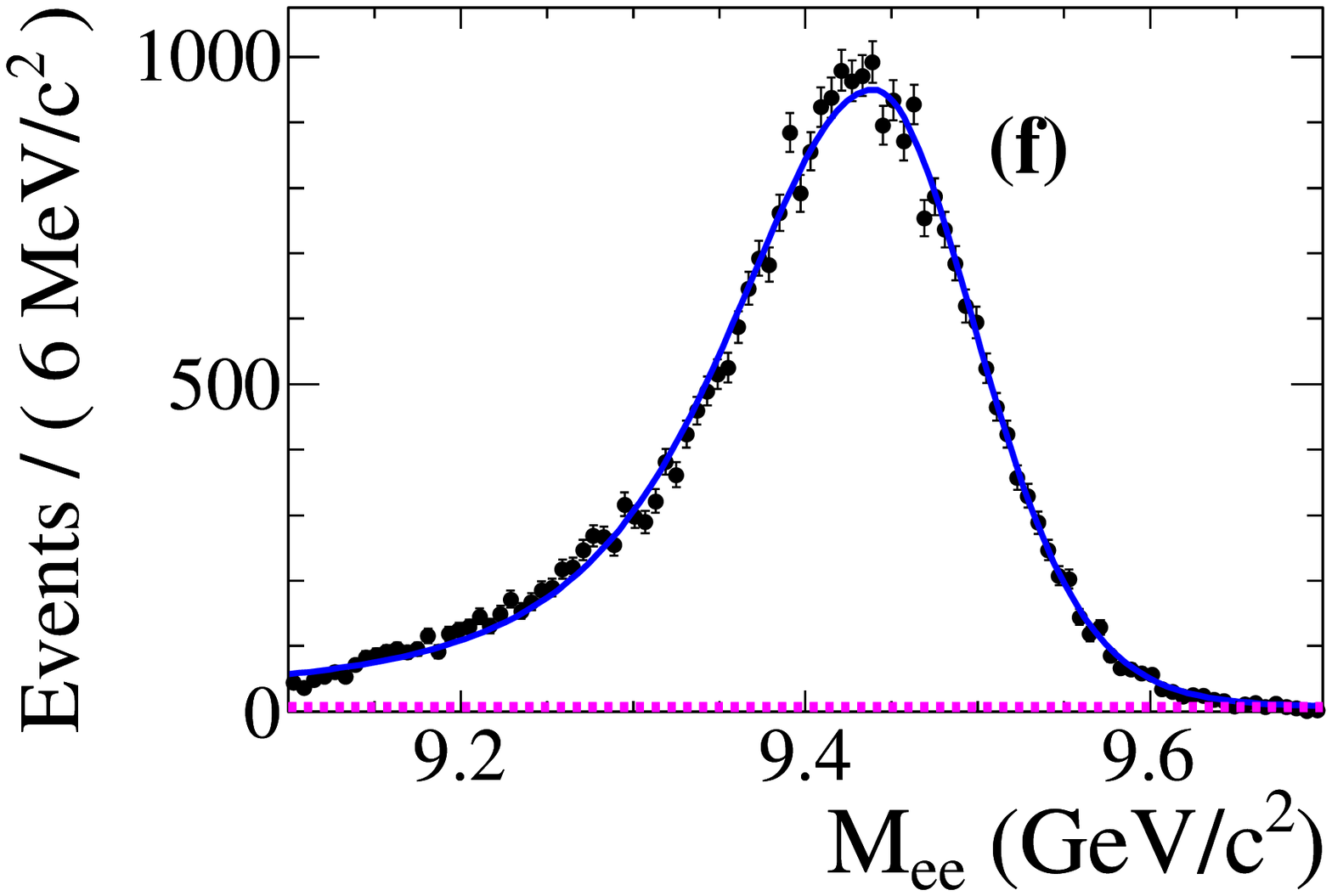}\\
\end{tabular}
\end{center}
\caption{$\Delta M_{\pi\pi}$ and $M_{\ell\ell}$ distributions for (a,b) 
$\TwoS\to\pipi\OneS\to\pipi\mumu$ candidates, (c,d) 
$\ThreeS\to\pipi\OneS\to\pipi\mumu$ candidates, and (e,f) 
$\ThreeS\to\pipi\OneS\to\pipi\epem$ candidates.
Data are represented by dots, the fit results as solid curves and the background components by the dashed curves.}
\label{fig:fit_dipion} 
\end{figure}

\begin{figure}[htdp]
\begin{center}
\begin{tabular}{ll}
\includegraphics[scale=0.23]{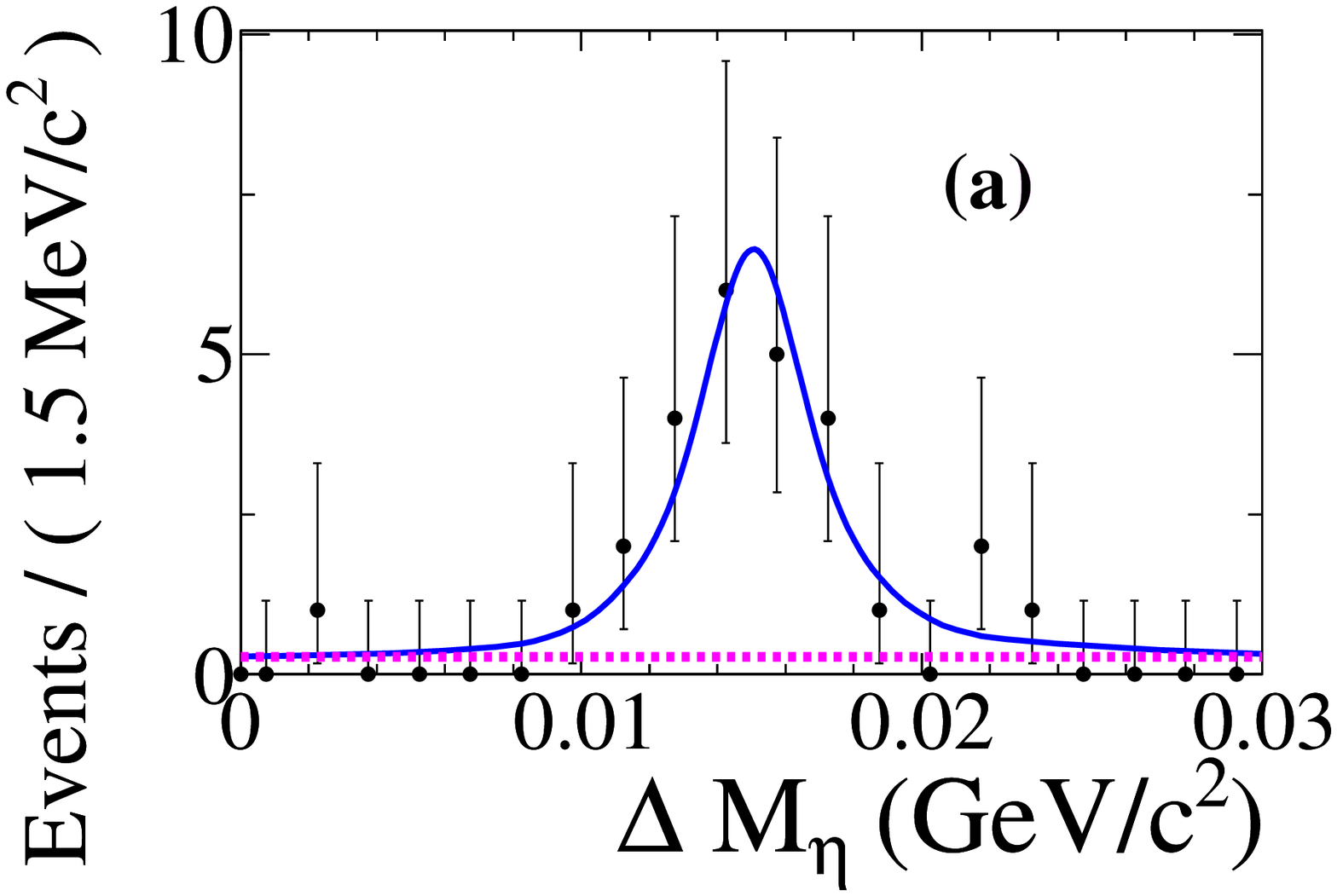}&
\includegraphics[scale=0.23]{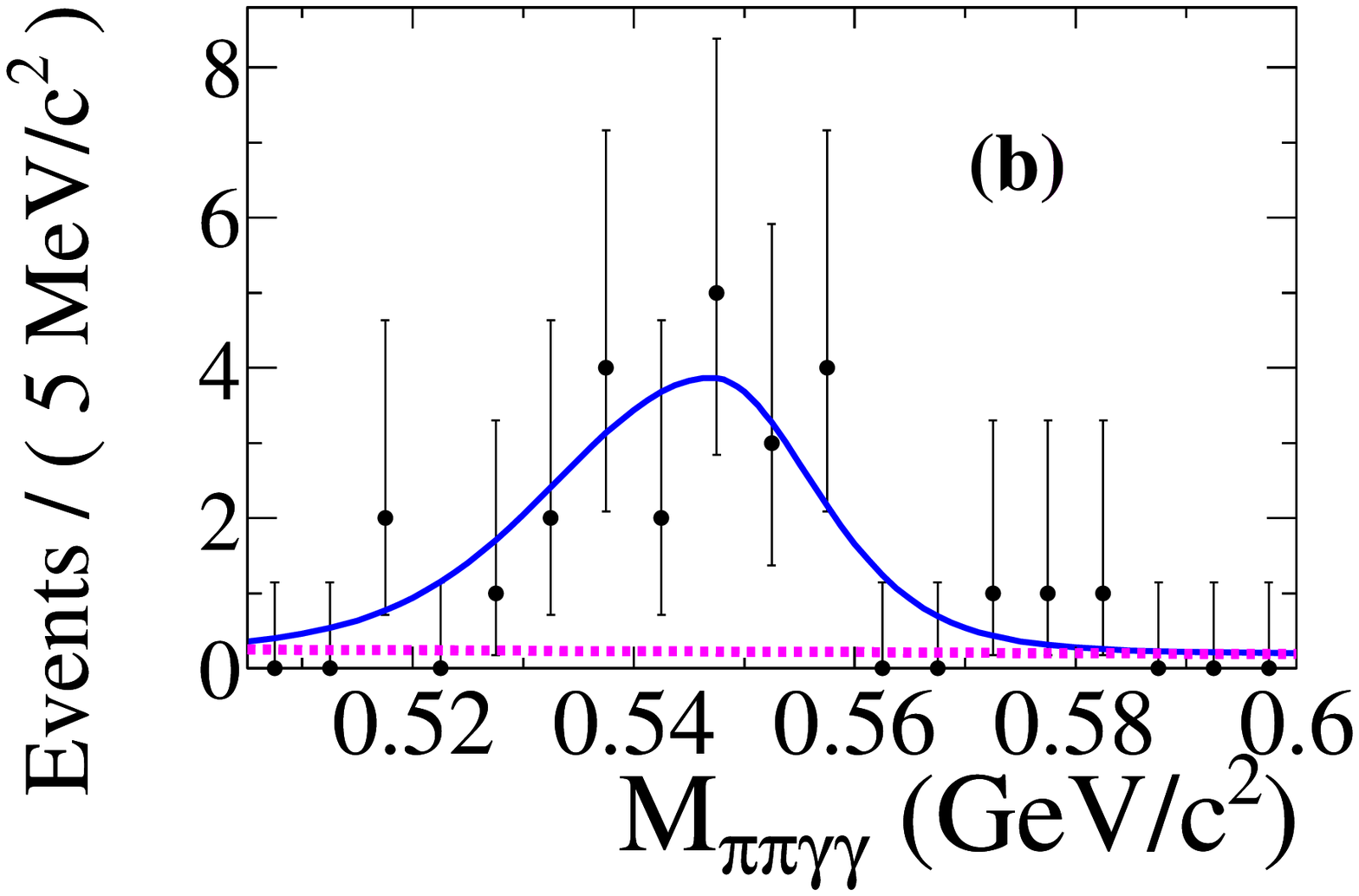}\\
\includegraphics[scale=0.23]{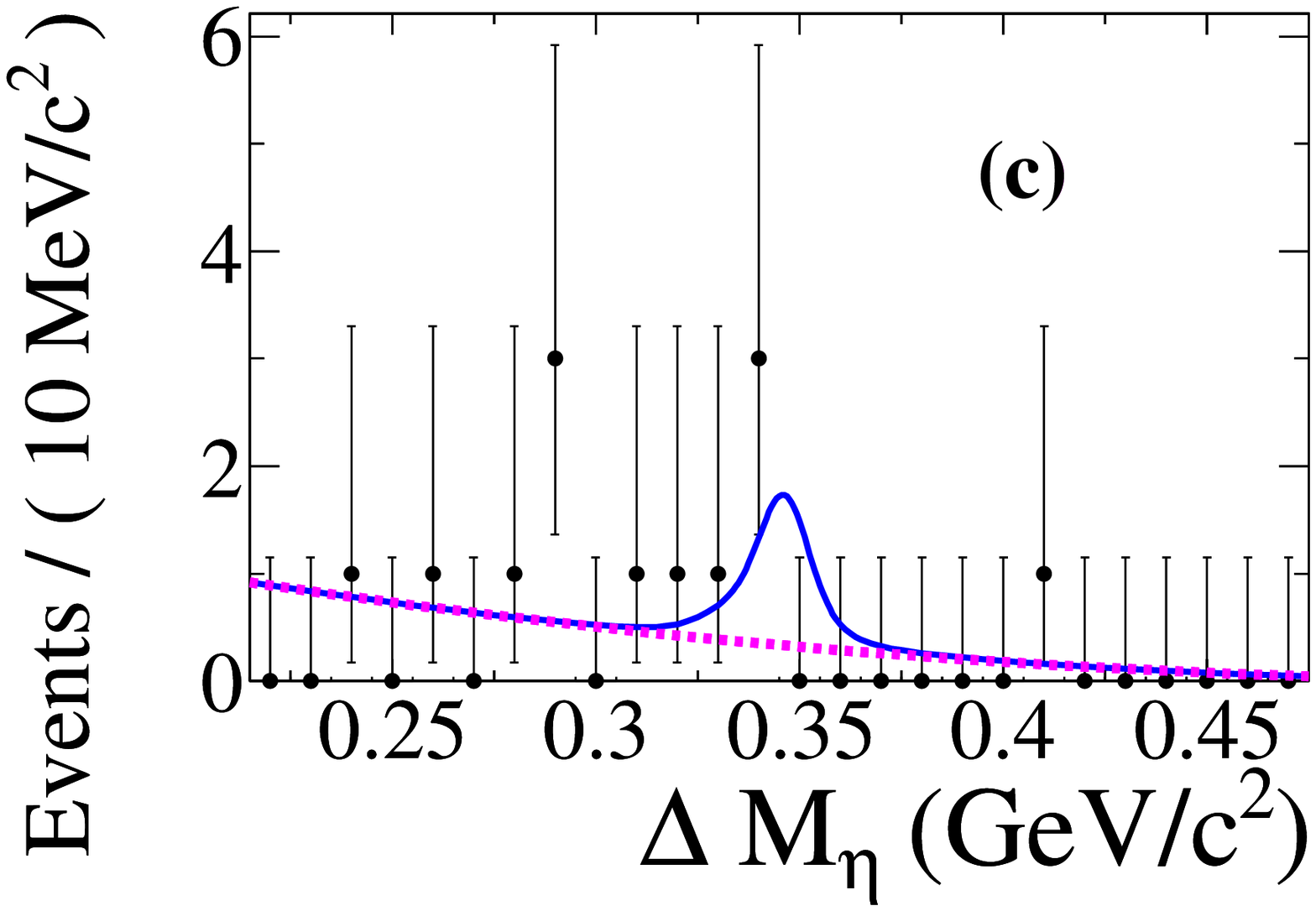}&
\includegraphics[scale=0.23]{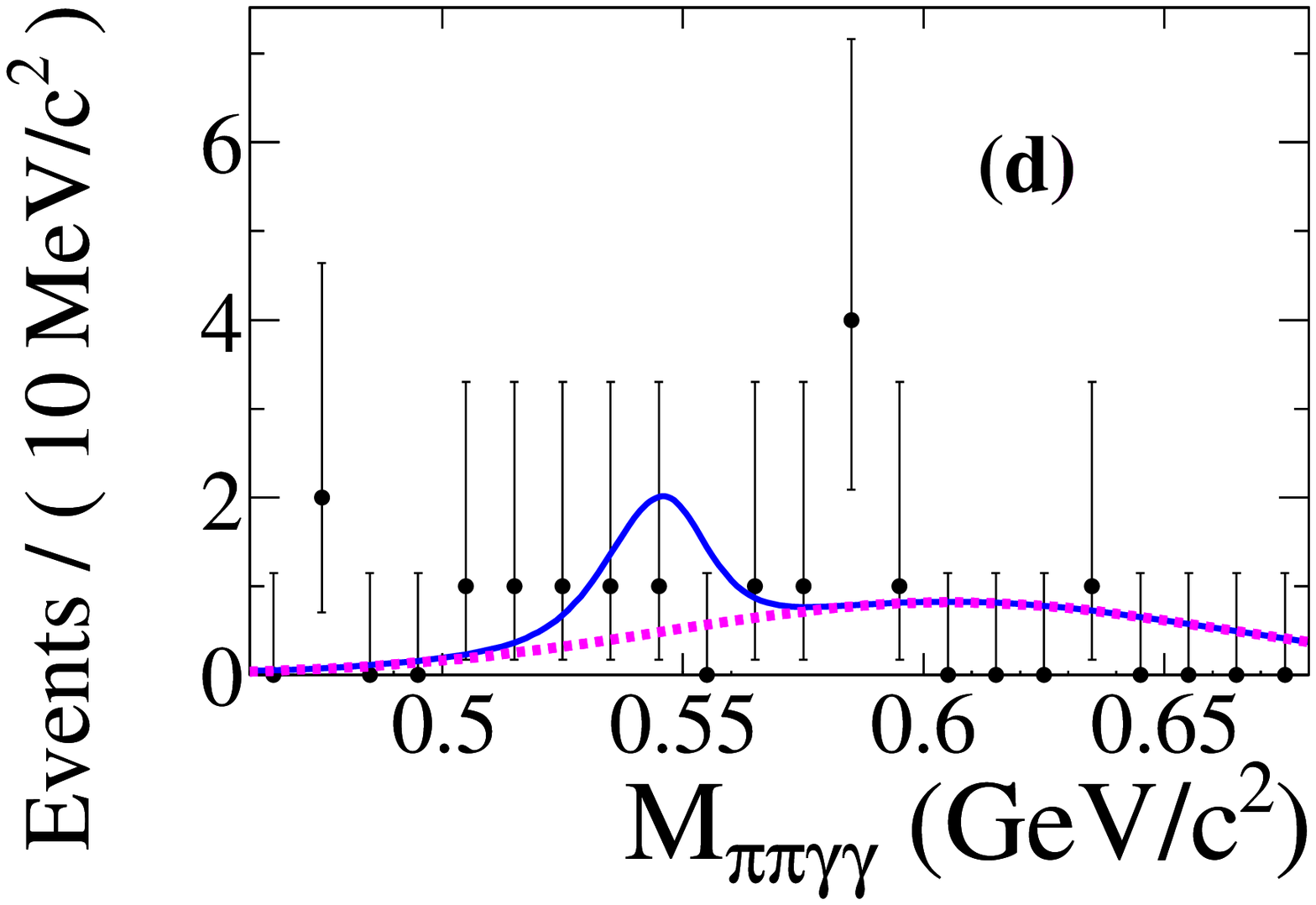}\\
\includegraphics[scale=0.23]{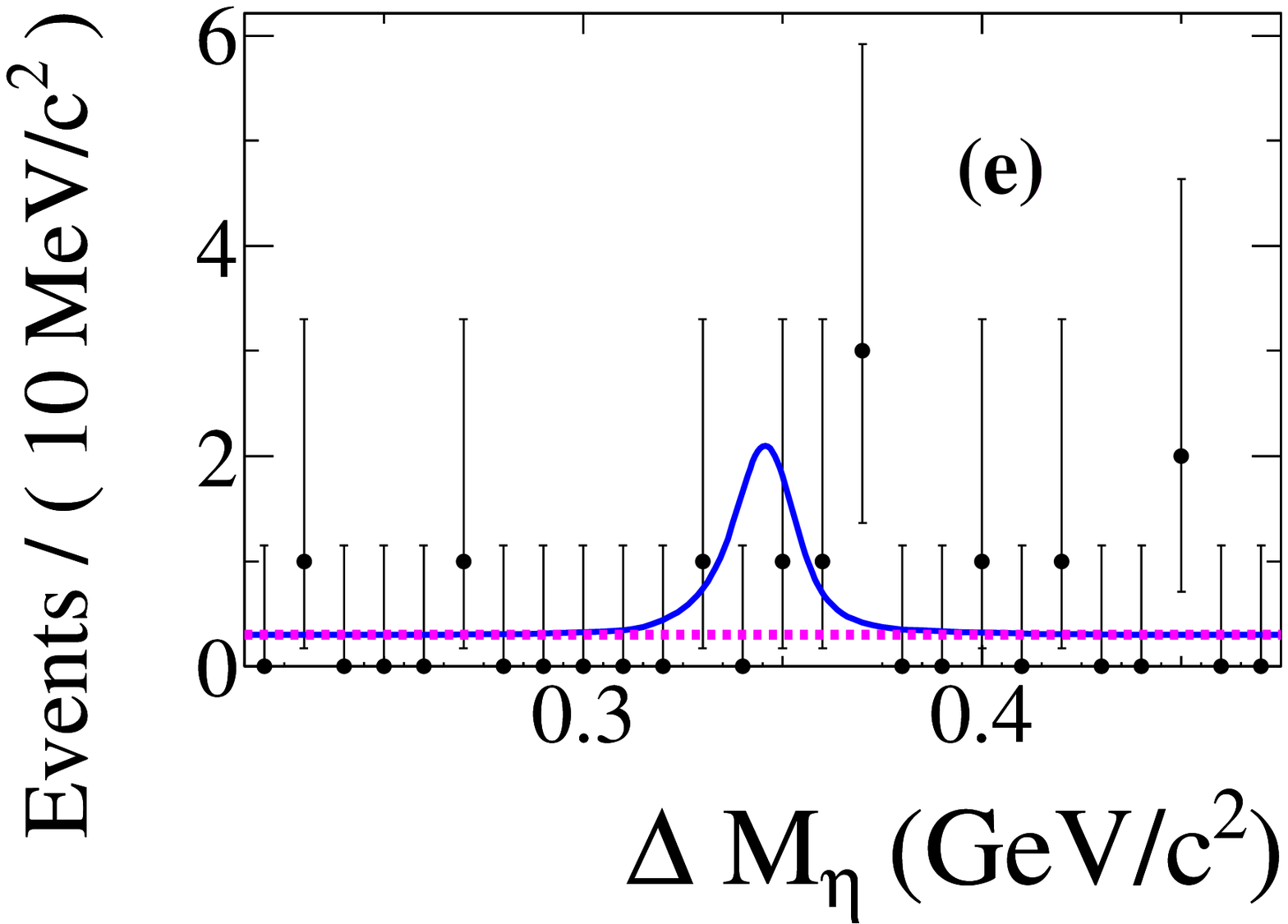}&
\includegraphics[scale=0.23]{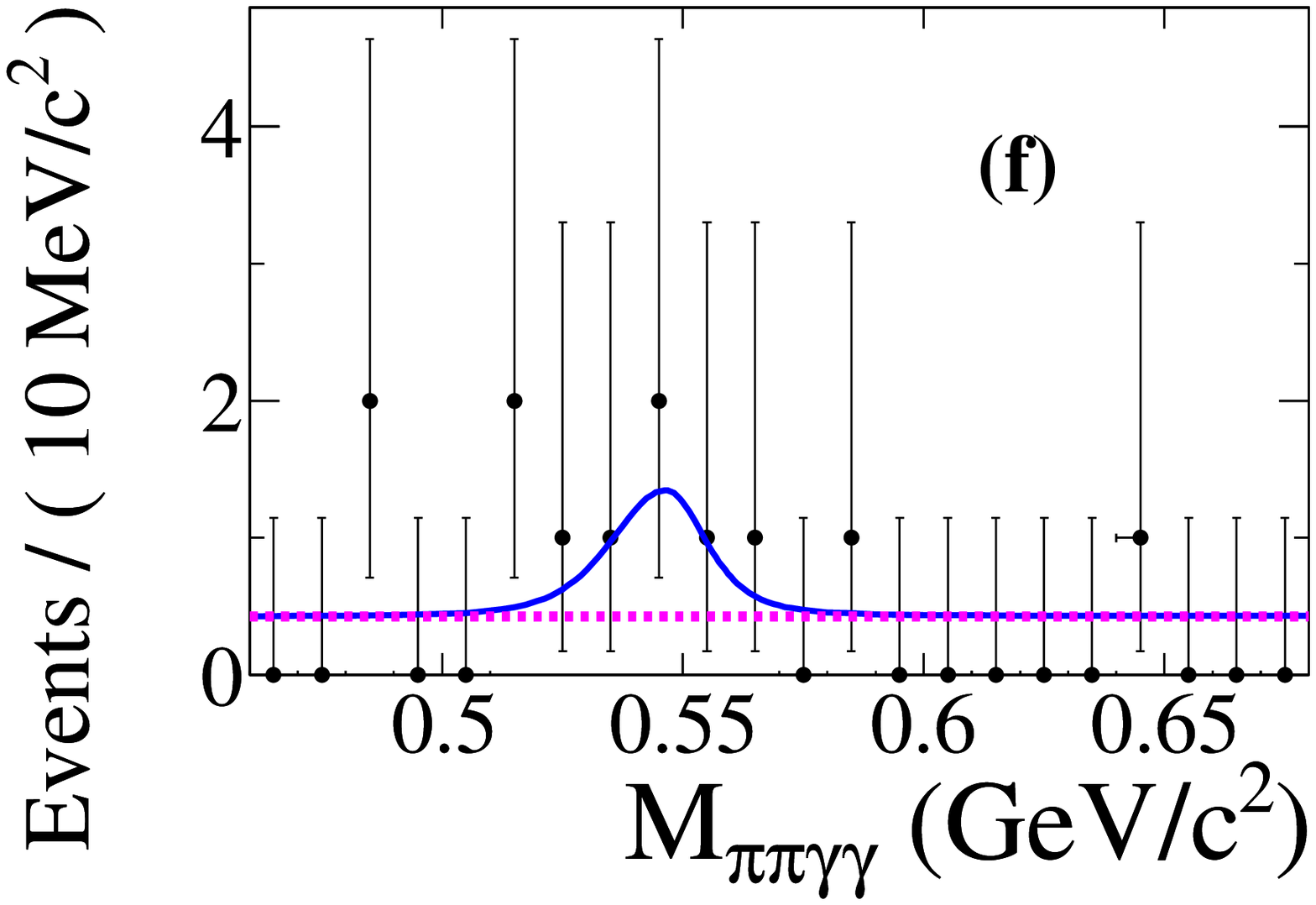}\\
\end{tabular}
\end{center}
\caption{$\Delta M_\eta$ and $M_{\pi\pi\gamma\gamma}$ distributions for (a,b) $\TwoS\to\eta\OneS\to\pipi\gamma\gamma\mumu$ candidates, (c,d) $\ThreeS\to\eta\OneS\to\pipi\gamma\gamma\mumu$ candidates, and (e,f) $\ThreeS\to\eta\OneS\to\pipi\gamma\gamma\epem$ candidates.
Data are represented by dots, the fit results as solid curves and the background components by the dashed curves.}
\label{fig:fit_eta3p} 
\end{figure}

\begin{figure}[htdp]
\begin{center}
\begin{tabular}{ll}
\includegraphics[scale=0.23]{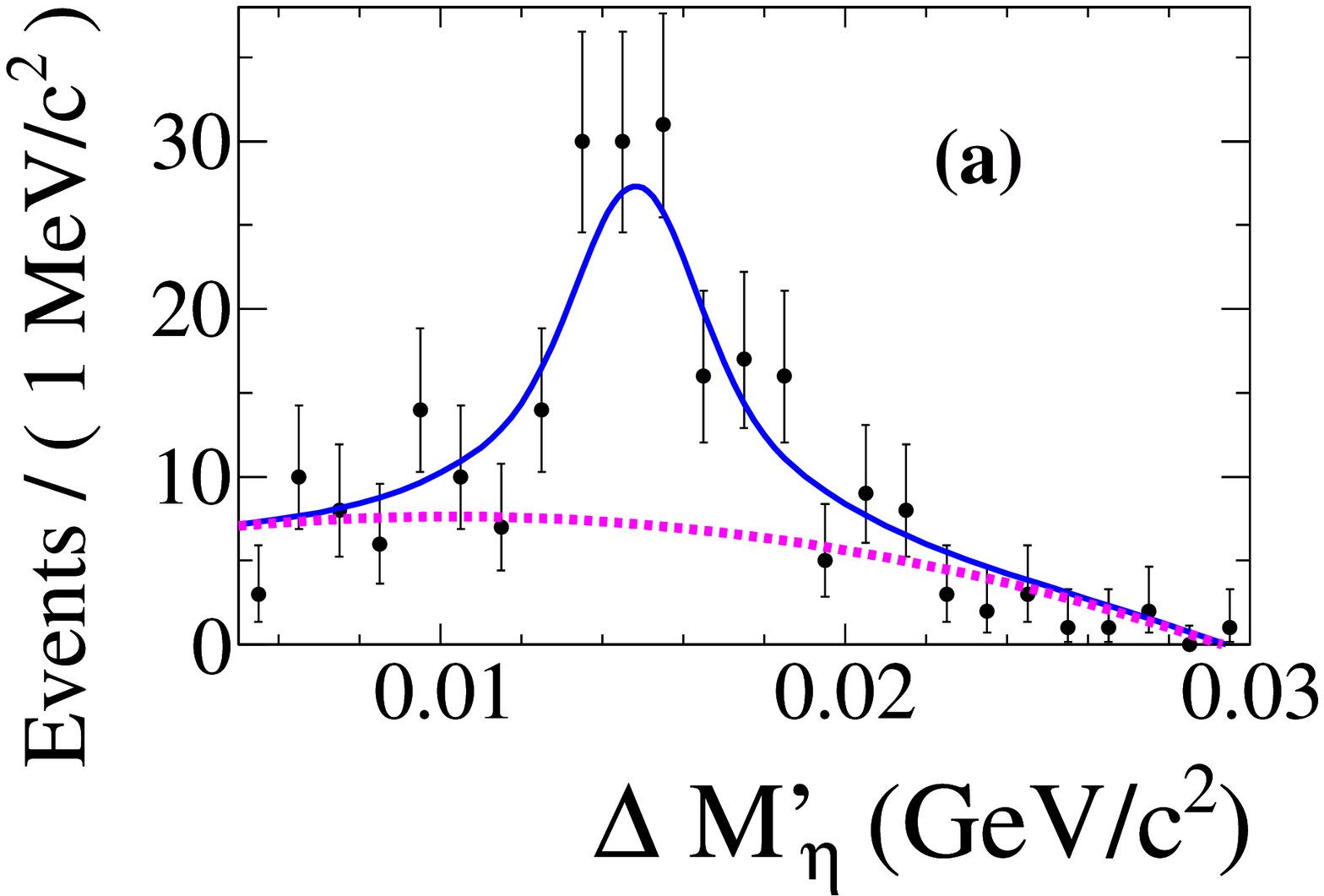}&
\includegraphics[scale=0.23]{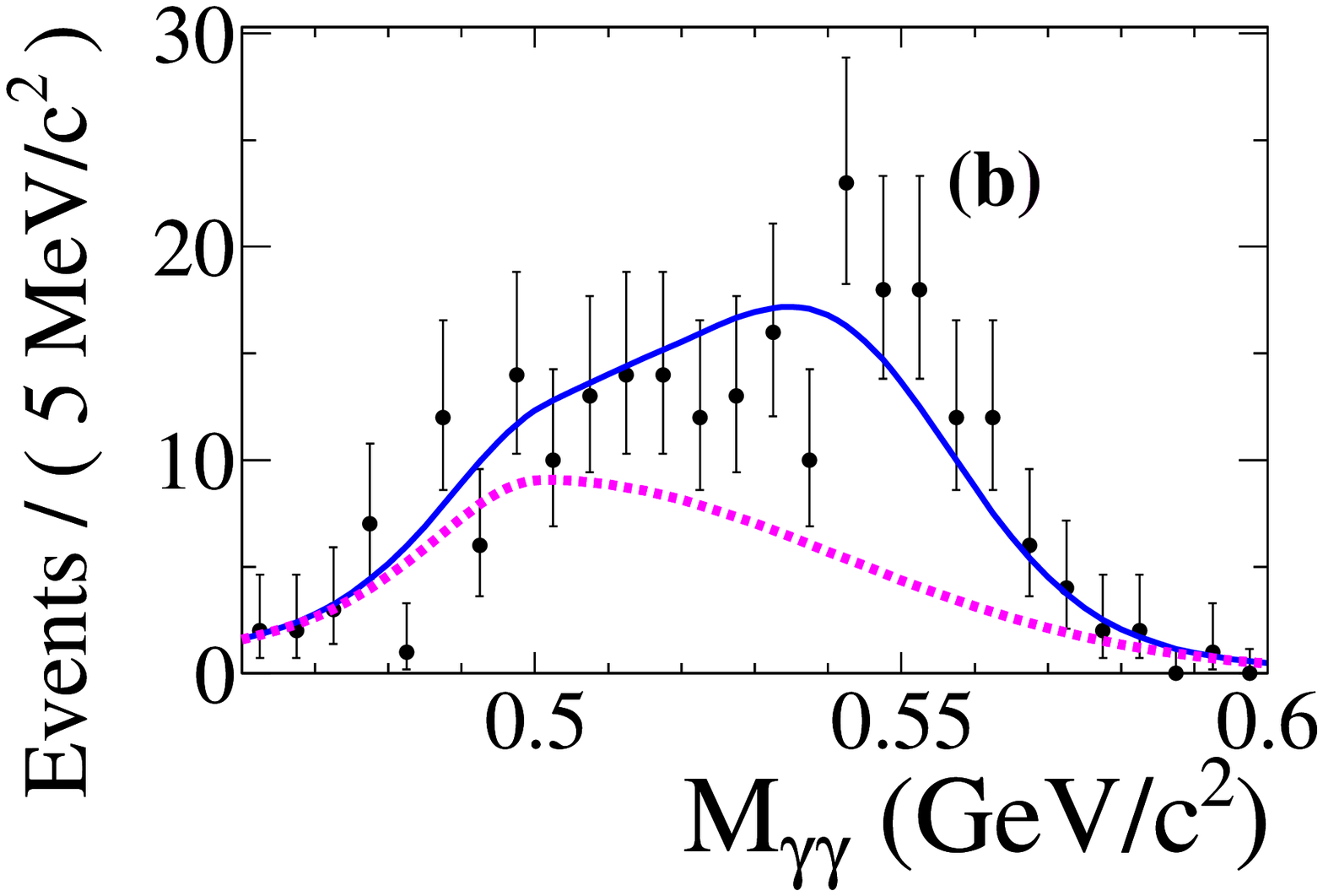}\\
\includegraphics[scale=0.23]{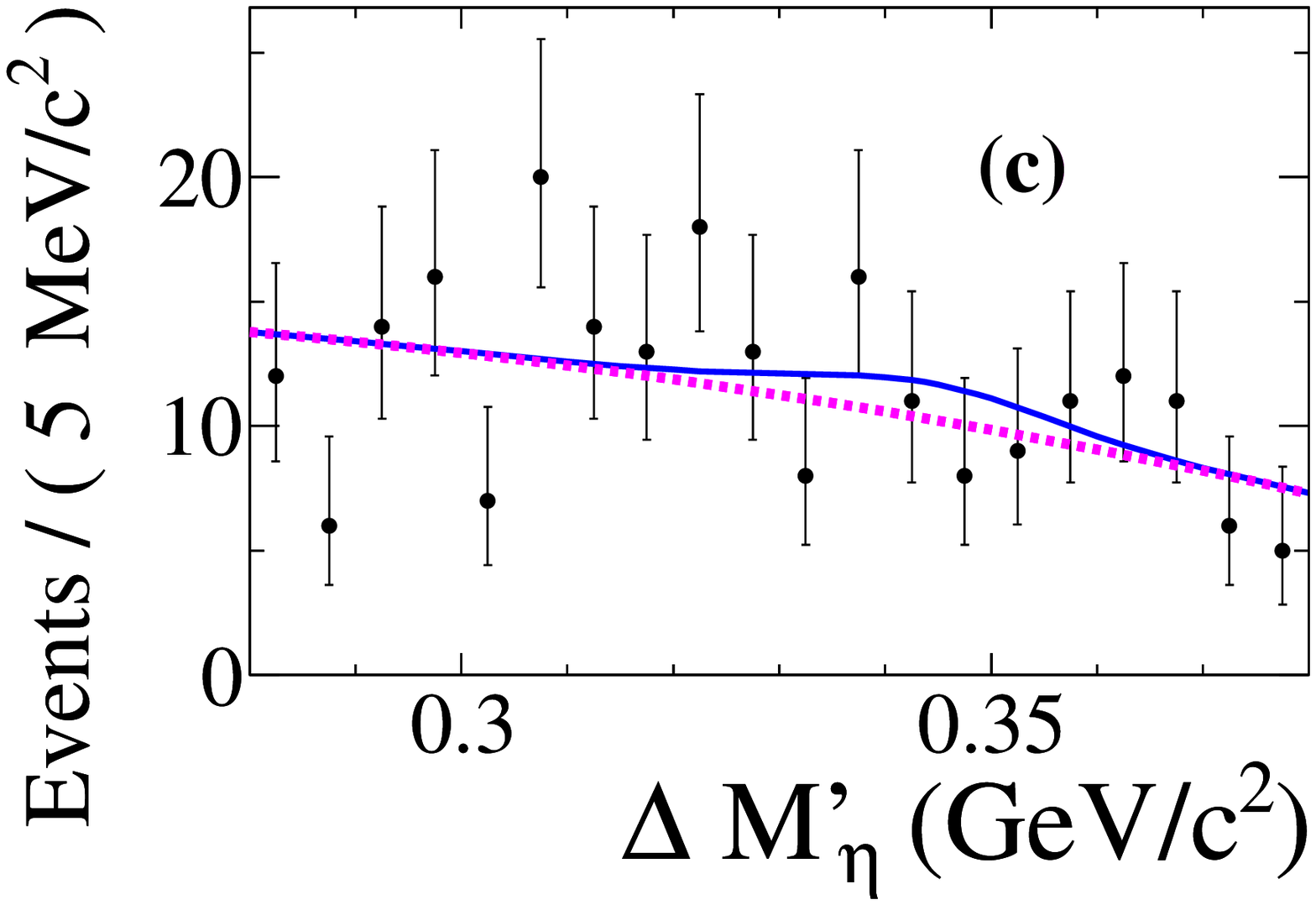}&
\includegraphics[scale=0.23]{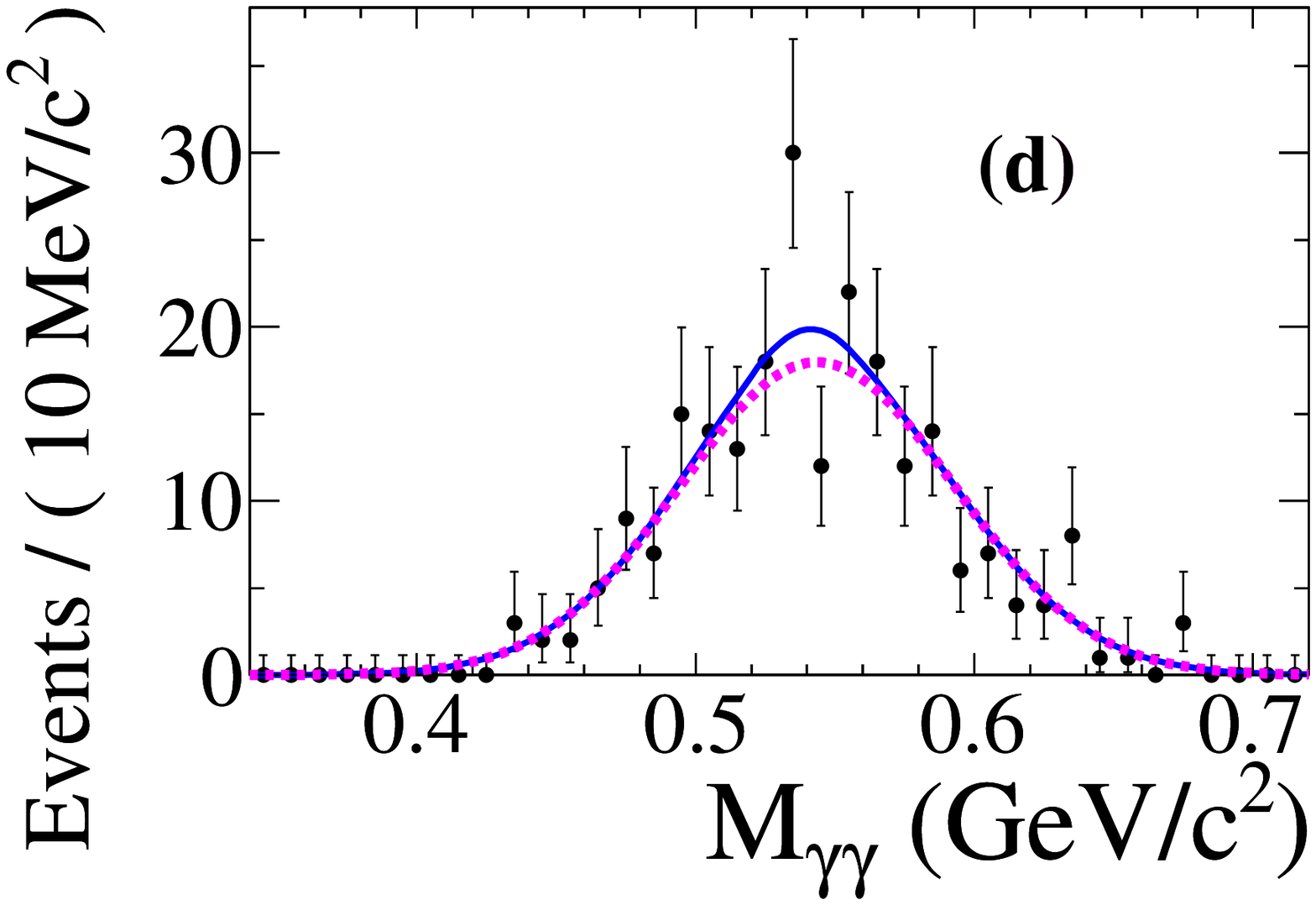}\\
\end{tabular}
\end{center}
\caption{$\Delta M'_\eta$ and $M_{\gamma\gamma}$ distributions for (a,b) $\TwoS\to\eta\OneS\to\gamma\gamma\mumu$ candidates, and (c,d) $\ThreeS\to\eta\OneS\to\gamma\gamma\mumu$ candidates.
Data are represented by dots, the fit results as solid curves and the background components by the dashed curves.}
\label{fig:fit_eta2g} 
\end{figure}

The signal yields are extracted with a 2-dimensional, unbinned, extended, maximum-likelihood fit to the measured distribution of a pair of variables. 
For the $\Upsilon(nS)\to\pipi\OneS$  transitions we fit the $\Delta M_{\pi\pi}$ versus $M_{\ell\ell}$ distribution, 
both calculated from the measured values prior to the invariant mass constraint (Fig.~\ref{fig:fit_dipion}). 
For the $\Upsilon(nS)\to\eta\OneS$ transitions with $\eta\to\pipi\pi^0$ decays, we fit the $\Delta M_\eta$ versus $M_{\pi\pi\gamma\gamma}$ distribution, where $\Delta M_\eta\equiv M_{\pi\pi\gamma\gamma \ell\ell}-M_{\ell\ell}-M_{\pi\pi\gamma\gamma}$ and $M_{\pi\pi\gamma\gamma}$ is the invariant mass of the $\eta$ decay products (Fig.~\ref{fig:fit_eta3p}).
For the $\Upsilon(nS)\to\eta\OneS$ transitions with $\eta\to\gamma\gamma$ decays, we fit the  $\Delta M'_\eta$ versus $M_{\gamma\gamma}$ distribution, where $\Delta M'_\eta\equiv M_{\gamma\gamma \ell\ell}-M_{\ell\ell}-M_{\gamma\gamma}$ (Fig.~\ref{fig:fit_eta2g}).

Each observed distribution is fit to a sum of a signal and a background component. The functional form of the probability density functions (PDFs) for signal and background have been determined from MC samples. The signal PDFs are described by double or triple Gaussian functions, or by a Gaussian-like analytical function with mean value $\mu$ but different widths, $\sigma_{L,R}$, on the left side (for $x<\mu$) and on the right side (for $x>\mu$) plus asymmetric tails $\alpha_{L,R}$, defined as:

\begin{equation}
\displaystyle{\cal F}(x) = \exp\Big\{ -\frac{(x-\mu)^2}{2\sigma^2_{L,R}+\alpha_{L,R}(x-\mu)^2} \Big\}. \label{eqn:Cruijff}
\end{equation}

The PDFs used to model the signal and background shapes in each fit are given in Table~\ref{tab:PDFs}. 
The free parameters in each fit are the signal and background yields and the parameter of the background PDFs of Table~\ref{tab:PDFs}. The signal shape parameters are also floated in the fits to the $\Upsilon(nS)\to\pipi\ellell$ samples, while they are fixed to the values determined from MC samples in all other cases.

\begin{table*}[htb]
\caption{Functions used to model the signal and background PDFs.}
\begin{center}
\begin{tabular}{l|ccc|ccc}
\hline
\hline
\noalign{\vskip1pt}
Final state & 1$^{st}$ Variable & Signal & Background & $2^{nd}$ Variable & Signal & Background\\
\hline
\noalign{\vskip1pt}
$\Upsilon(nS)\to\pipi\ellell $ & $\Delta M_{\pi\pi}$ & triple Gaussian & 0$^{th}$order poly  
                                          &$M_{\ell\ell}$ & Eq.(\ref{eqn:Cruijff}) & 0$^{th}$order poly\\
$\TwoS\to\pipi\gamma\gamma\mumu$ & $\Delta M_\eta$ & triple Gaussian & 0$^{th}$order poly 
						& $M_{\pi\pi\gamma\gamma}$ & Eq.(\ref{eqn:Cruijff}) & 1$^{st}$order poly \\
$\TwoS\to \gamma\gamma\mumu$ & $\Delta M'_\eta$ & triple Gaussian & 2$^{nd}$order poly 
						& $M_{\gamma\gamma}$ & double Gaussian & Eq.(\ref{eqn:Cruijff})\\
$\ThreeS\to\pipi\gamma\gamma\ellell$ & $\Delta M_\eta$ & triple Gaussian & 2$^{nd}$order poly 
						& $M_{\pi\pi\gamma\gamma}$ & Eq.(\ref{eqn:Cruijff}) & 2$^{nd}$order poly\\
$\ThreeS\to\gamma\gamma\mumu$ & $\Delta M'_\eta$ & double Gaussian & 1$^{st}$order poly 
						& $M_{\gamma\gamma}$ & double Gaussian & Gaussian\\
\hline
\hline						
\end{tabular}
\end{center}
\label{tab:PDFs}
\end{table*}

The number of signal candidates returned by the fits is reported in Table~\ref{tab:stat}. We estimate the signal significance in standard deviations as $\sqrt{2\log[{\cal L}(N)/{\cal L}(0)]}$, where $\displaystyle {\cal L}(N)/{\cal L}(0)$ is the ratio between the likelihood values for a fit that includes a signal yield $N$ and a fit with a background hypothesis only. For the $\TwoS\to\eta\OneS$ transition the signal significance is 8.0$\sigma$ for the $\pipi\gamma\gamma\mumu$ final states and 8.5$\sigma$ for the $\gamma\gamma\mumu$ ones. For the $\ThreeS\to\eta\OneS$ we find no evidence of a signal in any of the final states considered and calculate 90$\%$ CL upper limits on the number of signal events ($N_{UL}$) as $\int_0^{N_{UL}}{\cal L}(N)dN=0.9\times\int_0^\infty {\cal L}(N)dN$.
The efficiencies with which signal events satisfy the selection criteria ($\epsilon_{sel}$) are determined using simulated signal samples. Corrections are applied to account for differences between data and MC in lepton identification and $\piz$ reconstruction efficiencies. 
The corrected values are also reported in Table~\ref{tab:stat}.

\begin{table}[htdp]
\caption{Efficiencies ($\epsilon_{sel}$) and number of signal events ($N$) for each channel; upper limit at 90$\%$ CL ($N_{UL}$) is given in parentheses. Uncertainties are statistical only.}
\begin{center}
\begin{tabular}{lccc}
\hline
\hline
Transition                              & Final state                                &$\epsilon_{sel}$ (\%) & $N$\\
\hline
$\TwoS\to\pipi\OneS$           & $\pipi\mumu$                         & 39.1                            & 170061$\pm$413\\
\hline
\multirow{2}{*}{$\TwoS\to\eta\OneS$}            & $\pipi\gamma\gamma\mumu$  &   18.5   &22$\pm$5\\
                                                                              & $\gamma\gamma\mumu$         &    37.2  &90$\pm$14\\ 
\hline
\hline
\multirow{2}{*}{$\ThreeS\to\pipi\OneS$}        & $\pipi\epem$                   & 25.0    & 31330$\pm$186\\
                                                                             & $\pipi\mumu$                  & 42.8    & 58500$\pm$247\\
\hline
\multirow{3}{*}{$\ThreeS\to\eta\OneS$}         & $\pipi\gamma\gamma\epem$   & 18.1   & 4$\pm$2 ($<$8)\\
                                                                             & $\pipi\gamma\gamma\mumu$  &  8.9   &4$\pm$2 ($<$8)\\
                                                                             & $\gamma\gamma\mumu$         & 18.5   & 7$\pm$11 ($<$26)\\
\hline
\hline
\end{tabular}
\end{center}
\label{tab:stat}
\end{table}

Possible sources of systematic uncertainty are considered in addition to those on the number of $\Upsilon(nS)$ ($N_\Upsilon$) and on the  values for secondary branching fractions ($\BR_{secondary}$)~\cite{ref:PDG2010}.
The uncertainties on charged-particle track and single $\gamma$ or $\pi^0$ reconstruction efficiencies are determined by a comparison between data and MC events using independent control samples of $\tau$ pair events, each $\tau$ decaying to either one or three charged-particle tracks. 
The systematic uncertainty on the muon or electron identification probability is estimated by comparing the values determined in the $\Upsilon(nS)\to\pipi\OneS$ mode in data and MC samples. 
For each discriminating variable, we compare the distribution for the signal component  deconvolved from data with the maximum likelihood fit used for the extraction of the yields~\cite{ref:sPlot} to the distribution obtained in the MC.  The related systematic uncertainty is estimated as the change in event selection efficiency induced by the difference between the distributions.
The systematic uncertainties on the $\Delta M_{\pi\pi}$ and $E^*_{\gamma1,2}$ vetoes for cross-feed dipion and radiative transitions are estimated by comparing the corresponding efficiencies in data and MC samples.
In order to take into account possible discrepancies between simulation and data, the dipion events are generated using values for the transition matrix elements varied of $\pm1\sigma$ with respect to those measured by CLEO~\cite{ref:CLEO_dipion}. The difference in the efficiency is treated as a systematic uncertainty.
The systematic uncertainty due to the choice of signal and background PDFs is estimated by using different functions, or different values for the fixed parameters.
The complete list of contributions to the systematic uncertainty is summarized in Table~\ref{tab:syst}. The total systematic uncertainty for each dataset is estimated by summing all the contributions in quadrature.

\begin{table*}[htdp]
\caption{Sources of systematic uncertainty on the branching fractions $\BR$ and on the ratios of partial widths, for each channel analyzed. All errors are given in percent. When both of the leptonic $\OneS$ decays are analyzed, the values in parentheses refer to the corresponding $\epem$ final states.}
\begin{center}
\begin{tabular}{l|ccc|ccc}
\hline
\hline
            &   \multicolumn{3}{c|}{$\TwoS\to$}                                            & \multicolumn{3}{c}{$\ThreeS\to$} \\
\cline{2-7}
            &  $\pipi\OneS$ & \multicolumn{2}{c|}{$\eta\OneS$}                  & $\pipi\OneS$     & \multicolumn{2}{c}{$\eta\OneS$}\\
\hline
\noalign{\vskip1pt}
Source &                        & $\eta\to\pipi\pi^0$ &\hspace{0.5cm}  $\eta\to\gamma\gamma$ &                            & $\eta\to\pipi\pi^0$ &\hspace{0.5cm} $\eta\to\gamma\gamma$\\
\hline
$N_\Upsilon$    & \multicolumn{3}{c|}{0.9}                                                      & \multicolumn{3}{c}{1.0}\\
Tracking             & 1.4             & 1.4                        & 1.0                                     & 2.5                      &  2.5                               	   & 1.7\\
$\pi^0/\gamma$ & -                 & 3.6                        & 3.6                                    & -                           & 3.6                               	   & 3.6\\
Lepton identification           &  \multicolumn{3}{c|}{1.1}                                                     & 1.0 (1.2)              & 1.0 (1.2)                                  & 1.0\\
$\pipi$ model      & 0.5             &  -                           & -                                         & 0.4 (1.5)             &  -                                              & -   \\
Selection           & 0.4             &  2.6                      & 5.5                                      & 0.9 (1.2)              &  4.4 (5.3)                                 & 5.6  \\
PDFs                 & 0.1             &  5.4                      & 5.0                                      & 0.1                       & 5.4     			            & 5.0  \\
\hline
Total $\BR$      & 2.9            & 7.6                       & 8.7                                       & 3.6 (4.1)                & 8.6 (9.1)                		  & 8.1 \\
\hline
Total ratio        &                   & 7.2                       & 8.3                                      &                                & 8.3 (8.9)				  & 7.8 \\
\hline
\hline
\end{tabular}
\end{center}
\label{tab:syst}
\end{table*}

\begin{table*}[htdp]
\caption{Measured branching fractions and ratios of partial widths for hadronic $\Upsilon(nS)$ transitions. The first uncertainty is statistical, the second systematic. All ULs are at 90$\%$ of CL. The PDG values and the relevant predictions are given also.}
\begin{center}
\begin{tabular}{lr|c|c|c}
\hline
\hline
\multicolumn{2}{c|}{ } & This work                        & PDG~\cite{ref:PDG2010} & Predictions~\cite{ref:Kuang,ref:Simonov,ref:Meng}\\
\hline
\noalign{\vskip1pt}
$\BR[\TwoS\to\eta\OneS]$ & $(10^{-4})$   & ~~$2.39\pm0.31\pm0.14$   & $2.1^{+0.8}_{-0.7}$         &  7-16 \\
$\BR[\TwoS\to\pipi\OneS]$ & $(10^{-2})$  & $17.80\pm0.05\pm0.37$ & $18.1\pm0.4$                     & 40\\
$\displaystyle\frac{\Gamma[\TwoS\to\eta\OneS]}{\Gamma[\TwoS\to\pipi\OneS]}$ & $(10^{-3})$ & ~~$1.35\pm0.17\pm0.08$ & $1.2\pm0.4$ & 1.7-3.8\\
\noalign{\vskip1pt}
\hline
\noalign{\vskip1pt}
$\BR[\ThreeS\to\eta\OneS]$ & $(10^{-4})$ & $<1.0$                             & $<1.8$                                &  5-10\\
$\BR[\ThreeS\to\pipi\OneS]$ & $(10^{-2})$                  & ~~$4.32\pm0.07\pm0.13$   & $4.40\pm0.10$                   & 5\\
$\displaystyle\frac{\Gamma[\ThreeS\to\eta\OneS]}{\Gamma[\ThreeS\to\pipi\OneS]}$ & $(10^{-3})$ & $<2.3$ & $<4.2$ & 11-20\\
\noalign{\vskip1pt}
\hline
\hline
\end{tabular}
\end{center}
\label{tab:res}
\end{table*}

The value of the branching fraction  (${\cal B}$), or upper limit on the branching fraction, for each mode is: 
\begin{equation}
\displaystyle {\cal B}= \frac{N}{\epsilon_{sel}\times N_{\Upsilon}\times\BR_{secondary}},
\end{equation}
where $N$ is the signal yield or upper limit on the signal yield.
For a given channel, when both the leptonic $\OneS$ decays are available, their signal yields are first combined in a weighted average, where the weight is the inverse of the squared sum of the statistical and the systematic uncertainties on each yield, considering only the systematic contributions that are uncorrelated. We assume ${\cal B}[\OneS\to\epem]={\cal B}[\OneS\to\mumu]$~\cite{ref:PDG2010}.
For the $\eta$ transitions, the signal yields extracted from the two different $\eta$ decays are combined with the same weighted average technique.
The results are shown in Table~\ref{tab:res}.

We can also provide improved measurements of the differences between the $\Upsilon$ invariant masses, using the fitted value of $\Delta M_{\pi\pi}$ for both the $\TwoS\to\pipi\OneS$ and the $\ThreeS\to\pipi\OneS$ transitions. The values are $562.170\pm0.007{\rm(stat.)}\pm0.088{\rm(syst.)}\mevcc$ and $893.813\pm0.015{\rm(stat.)}\pm0.107{\rm(syst.)}\mevcc$, respectively, where the latter value is obtained as a weighted average of the values for the electron and muon samples. The systematic uncertainties are due mainly to the track momentum measurement, which is related to the knowledge of the amount of detector material and of the magnetic field~\cite{ref:Babar_taumass}.

We have presented a study of $\Upsilon(3S)\to\OneS$ and $\Upsilon(2S)\to\OneS$ hadronic transitions. We have reported an improved measurement of ${\cal B}[\TwoS\to\eta\OneS]$ and a 90\% CL  UL on  ${\cal B}[\ThreeS\to\eta\OneS]$ compatible with, and more precise than, earlier measurements~\cite{ref:CLEO_eta}, thus, further constraining theoretical predictions (see Table~\ref{tab:res}). We have also presented new measurements of ${\cal B}[\Upsilon(nS)\to\pipi\OneS]$ with $n=3,2$, which we find to be compatible with earlier measurements~\cite{ref:PDG2010}. 
Using the independent \babar\ measurement of ${\cal B}[\ThreeS\to X\TwoS]\times{\cal B}[\TwoS\to\pipi\OneS]$ in the inclusive dipion spectrum~\cite{ref:Babar_dipionNew}, we extract the value ${\cal B}[\ThreeS\to X\TwoS]=(10.0\pm0.6)\%$.

Improved measurements of the ratios $\Gamma[\Upsilon(nS)\to\eta\OneS]/\Gamma[\Upsilon(nS)\to\pipi\OneS]$, for which systematic uncertainties partially cancel, have been presented also ~\cite{ref:PDG2010}. The suppression of the $\Upsilon(nS)\to\eta\OneS$ transitions with respect to the $\Upsilon(nS)\to\pipi\OneS$ ones is confirmed to be higher than predicted by the QCDME ~\cite{ref:Kuang} and not compatible with other models~\cite{ref:Simonov, ref:Meng}.

We are grateful for the excellent luminosity and machine conditions
provided by our \pep2\ colleagues, 
and for the substantial dedicated effort from
the computing organizations that support \babar.
The collaborating institutions wish to thank 
SLAC for its support and kind hospitality. 
This work is supported by
DOE
and NSF (USA),
NSERC (Canada),
CEA and
CNRS-IN2P3
(France),
BMBF and DFG
(Germany),
INFN (Italy),
FOM (The Netherlands),
NFR (Norway),
MES (Russia),
MICIIN (Spain),
STFC (United Kingdom). 
Individuals have received support from the
Marie Curie EIF (European Union),
the A.~P.~Sloan Foundation (USA)
and the Binational Science Foundation (USA-Israel).

\end{document}